\documentclass[%
reprint,
nofootinbib,
amsmath,assume,
aps,
]{revtex4-2}

\usepackage[utf8]{inputenc}
\usepackage[table,dvipsnames]{xcolor}

\usepackage{caption}
\usepackage{subcaption}
\usepackage{graphicx}

\usepackage{dcolumn}
\usepackage{bm}
\usepackage{physics}
\usepackage[colorlinks=true,linkcolor=blue,citecolor=blue,urlcolor=
blue]{hyperref}
\usepackage{latexsym}

\usepackage{lipsum}
\usepackage{mwe}
\usepackage{multirow}
\usepackage{amsmath, amsthm, amssymb}

\newcommand{\munu}{\mu\nu}

\newcommand{\lrb}[1]{\left( #1 \right)}
\newcommand{\lrrb}[1]{\left[ #1 \right]}
\newcommand{\lrcb}[1]{\left\{ #1 \right\}}

\begin{document}
	
	\preprint{APS/123-QED}
	
	\title{The Convergence Problem Of Gradient Expansion \\ In The Relaxation Time Approximation } 

	\author{Reghukrishnan Gangadharan }
    \email{reghukrishnang@niser.ac.in }
    \affiliation{School of Physical Sciences, National Institute of Science Education and Research, An OCC of Homi Bhabha National Institute, Jatni-752050, India}
	
    \author{Victor Roy}
    \email{victor@niser.ac.in}
    \affiliation{School of Physical Sciences, National Institute of Science Education and Research, An OCC of Homi Bhabha National Institute, Jatni-752050, India}

	\date{\today}
	
\begin{abstract}
  
We obtain a formal integral solution to the 3+1 D Boltzmann Equation in relaxation time approximation. The gradient series obtained from this integral solution contains exponentially decaying non-hydrodynamic terms. It is shown that this gradient expansion can have a finite radius of convergence under certain assumptions of analyticity. We then argue that, in the relaxation time model, proximity to local thermal equilibrium is not necessary for the system to be described by hydrodynamic equations. 
\end{abstract}
	
	\maketitle
\textbf{Introduction:}
Hydrodynamics is an effective field theory describing near-equilibrium phenomena of many body systems. It finds application in a wide range of fields including high energy nuclear physics~\cite{Florkowski:2017olj}, astrophysics~\cite{PhysRevLett.122.221602,PhysRevD.95.083005,PhysRevLett.120.041101,PhysRevD.107.103031}, general relativity~\cite{Bhattacharyya:2007vjd,Rudolf_Baier_2008} and condensed matter physics \cite{annurev:/content/journals/10.1146/annurev-conmatphys-040521-042014,2017PhRvB..96b0401M, PhysRevB.104.184414}. Hydrodynamics is formulated by expanding the conserved currents like energy-momentum tensor, charge density, etc, in terms of the gradients of thermodynamic fields, temperature, chemical potential, and fluid flow velocity. Such an expansion is carried out assuming that the system under consideration is close to local thermal equilibrium and that the gradients of the thermodynamic quantities are small, i.e., when the mean free path is much smaller than the system size. However, as the gradient expansion is known to be divergent, adding higher-order gradient terms does not lead to a more accurate description of the exact dynamics. The factorial increase in the number of terms in each order of the gradient expansion was shown to imply a zero radius of convergence~\cite{PhysRevLett.110.211602,Buchel:2016cbj,Denicol:2016bjh, Heller:2015dha}. In the non-relativistic case, it was shown~\cite{10.1063/1.1706716}that the gradient expansion can only be considered valid as an asymptotic series.  

Despite the restricted domain of validity, hydrodynamics simulations have successfully explained a multitude of experimental data regarding flow and invariant yields of hadrons in relativistic heavy-ion experiments with a very short initialization time~\cite{Luzum:2009sb,Heinz:2013th,Teaney:2001av,Teaney:2000cw}. It is a longstanding puzzle of how the system reaches local thermal equilibrium in such a short time, given that the gradients are large and the partonic interactions are initially weak due to high temperature. The question becomes more pressing for smaller systems such as high multiplicity proton-proton collisions where hydrodynamics-like behaviour is also seen~\cite{CMS:2010ifv,ATLAS:2015hzw,CMS:2015fgy}. In order to explain this anomaly, several recent studies proposed that hydrodynamics could have a broader range of validity~\cite{Romatschke:2017acs,Romatschke:2017vte,Denicol:2019lio,Blaizot:2017ucy,Strickland:2018ayk,Dash:2020zqx}.

Extention of the applicability of hydrodynamics to non-equilibrium systems demands both a justification of the gradient expansion and a resolution of its non-convergence. For symmetries relevant to heavy ion collisions, conformal symmetry, and boost invariance, solutions to hydrodynamic equations showed early time attractor behavior, i.e., an initial non-equilibrium state showed hydrodynamic evolution even at large gradients. This behaviour was also reproduced for solutions to the Boltzmann equation~\cite{Florkowski:2013lza, Florkowski:2013lya,Denicol:2014tha,Heinz:2015gka}. However, it was shown in later studies that when conformal symmetry was broken, early-time attractors were no longer present for all variables of interest~\cite{Chattopadhyay:2021ive,Jaiswal:2022mdk,Jaiswal:2022udf}. 

The problem of the divergence of the gradient expansion remains even in highly symmetric systems~\cite{Denicol:2016bjh, Heller:2021oxl,Heller:2016rtz}. The asymptotic nature of the gradient expansion means that it does not capture the exponentially decaying transient modes~\cite{Aniceto:2015mto, Heller:2018qvh,Heller:2016rtz} of the exact solution. These terms become crucial in capturing the non-equilibrium behaviour of the exact solution and rendering the series expansion convergent. In several studies~\cite{Heller:2015dha,Heller:2018qvh, Basar:2015ava, Buchel:2016cbj} re-summation techniques were employed to obtain the transient terms from the gradient expansion. Though the re-summation results provide valuable insights, a comprehensive understanding of the hydrodynamisation process requires the study of the underlying microscopic theory and its macroscopic limit. 

The relativistic Boltzmann equation is a valuable tool in studying non-equilibrium phenomena in the weak coupling limit. It has been successfully used to derive causal theories of relativistic hydrodynamics~\cite{Denicol:2014xca, Denicol:2010xn,Denicol:2012cn,Jaiswal:2012qm,Jaiswal:2013npa,Jaiswal:2013vta,Jaiswal:2015mxa,Bhadury:2020puc,Dash:2021ibx,Panda:2020zhr,Panda:2021pvq}, and is used as a benchmark for such theories in reproducing non-equilibrium dynamics. Deriving the transient terms from the exact solution to the Boltzmann equation should validate the limiting behaviour obtained from re-summation techniques. For finite orders in gradient expansion, it was shown in~\cite{McNelis:2020jrn} that a transient structure can be obtained for the Boltzmann equation in the relaxation time model. It was conjectured from this that the exact solution to the relaxation time Boltzmann equation is the Borel re-summation of the gradient expansion. In this work, we obtain a formal 3+1 D solution to the RTA Boltzmann equation and rigorously prove the conjecture in~\cite{McNelis:2020jrn}. We show that under analyticity assumptions, the full gradient expansion is convergent. We interpret the exponentially decaying transient terms as free streaming contributions from the initial conditions. Further, we argue that a hydrodynamics description is valid even at large gradients as long as the transient terms are sufficiently small. 

 Throughout this text, we use natural units where $c=\hslash =k_{B} = 1$.

 \textbf{The Boltzmann Equation:}
The single particle phase space distribution function $f(x^{\mu},p^{\mu})$ is the probability density of finding a particle at position $x^{\mu}$ with momentum $p^{\mu}$. The Boltzmann equation 
 
 \begin{align}
     \lrb{p^{\mu}\pdv{}{x^{\mu}} + mK^{\mu}\pdv{}{p^{\mu}} - \Gamma^{\sigma}_{\munu}p^{\mu}p^{\nu}\pdv{}{p^{\sigma}}}f = C(f,f)\,, \label{Eq:BolEq}
 \end{align}
 determines the dynamical evolution of the distribution function in the presence of external forces~\cite{Cercignani:2002rbe}.Here $K^{\mu}$ is a 4-force and $\Gamma^{\sigma}_{\munu}$ are the Christoffel symbols. The four momenta $p^{\mu}$ satisfy the on-shell condition $g_{\munu}p^{\mu}p^{\nu} = m^2$ and the 4-force satisfy $g_{\munu}p^{\mu}K^{\mu}/p^{0} = 0$. Here $C(f,f)$ is the collision kernel. In this paper, we use the simplified Anderson-Witting relaxation time approximation(RTA)~\cite{Cercignani:2002rbe} kernel,
\begin{align}
    C(f,f) = -\frac{U\cdot p}{\tau_R}(f-f_{\text{eq}})\,.
\end{align}
The effective local equilibrium distribution function $f_{eq}$ is defined as,
\begin{equation}
    f_{\text{eq}} = \frac{1}{e^{\lrb{U\cdot p - \alpha}/T} + r},
\end{equation}
where $r$ takes values ${-1,0,1} $ corresponding to Bose-Einstein, Maxwell-Juttner, and Fermi-Dirac statistics respectively. The values of $U^{\mu}$ and $T$ are constructed using Landau-Matching conditions. $\tau_R$ is called the relaxation time and is in general a function of position and momenta.


\textbf{The Formal Solution:} In this section, we outline the derivation of the exact solution. For the sake of simplicity, we will be working in Minkowski space with $g_{\munu} = (1,-1,-1,-1)$ and with the force-free version of the Boltzmann equation, 
\begin{equation}\label{Eq:RtaMikBol}
    p^{\mu}\partial_{\mu}{f} = -U\cdot p\frac{(f-f_{\text{eq}})}{\tau_R}.
\end{equation}
The derivation is quite general and can be easily extended to arbitrary coordinates and with external forces with minor changes. The essential qualitative features however remain unchanged. For most of this paper, we suppress the momentum dependence which is assumed to be implied. 

To find a formal solution for the $3+1$ D Boltzmann equation, we employ the standard techniques of solving first-order partial differential equations using characteristic curves. The characteristic equations and the corresponding characteristic curves for Eq.\eqref{Eq:RtaMikBol} are
\begin{align}
    \dv{x^{\mu}}{s} =p^{\mu} &&\Rightarrow &&x^{\mu} = p^{\mu}s + c^{\mu}\,.
\end{align}
$s$ is the variable that parametrises the characteristic curve and $c^{\mu}$ is the integration constant representing the initial conditions. It is more insightful if we use $x^{0} = t$ as the parameter using the relations,
\begin{align}
    t &= p^{0} s +t_{0}  &&\Rightarrow &&  s = \frac{(t-t_{0})}{p^{0}}, \\
    x^{i} &= \frac{p^{i}}{p^{0}}(t-t_{0}) + c^{i} &&\Rightarrow &&  c^{i} = x^{i} - \frac{p^{i}}{p^{0}}(t-t_{0}). 
\end{align}
Note that the characteristic curves are the free particle trajectories. In the general case of curved spacetime and external force, the characteristic curve will be the particle trajectory under these external effects.

Given some initial condition on $x^{\mu}(t') = c^{\mu}$, any function $h(t)$ on the characteristic curve has the parametric dependence:
\begin{align}
    h(t,t') \Rightarrow h(x^{\mu} - \frac{p^{\mu}}{p^{0}}(t-t') )\,.
\end{align}
From here on we will assume this dependence is implied unless otherwise specified. Eq.\eqref{Eq:RtaMikBol} on the characteristic curve then reduces to the one-parameter form
\begin{equation}
    \dv{f(t,t')}{t} = -U(t,t')\cdot p\frac{(f(t,t')-f_{eq}(t,t'))}{\tau_R(t,t')}\,.
\end{equation}
and can be integrated to give the formal solution \eqref{Ap:ForSOl}
\begin{equation}\label{Eq:ExtSol}
    f(t) = e^{-\xi(t,t_{0})} f_{0}(t,t_0) + \int_{t_0}^{t}\frac{dt'}{p^0} \frac{U(t,t')\cdot p}{\tau_R(t,t')}e^{-\xi(t,t')}f_{eq}(t,t') \,,
\end{equation}
where $f_{0}$ is the initial distribution function and 
\begin{equation}
    \xi(t,t') = \int_{t'}^{t}\frac{U(t,t'')\cdot p}{\tau_R(t,t'')} \frac{dt''}{p^0}\,.
\end{equation}
The first term on the right-hand side of Eq.\eqref{Eq:ExtSol} is the free streaming initial distribution scaled by an exponentially decaying damping factor. Following~\cite{McNelis:2020jrn}, we will call the second term  in Eq.\eqref{Eq:ExtSol},
\begin{equation}\label{Eq:HydGen}
    f_{G}(x,t) = \int_{t_0}^{t}\frac{dt'}{p^0} \frac{U(t,t')\cdot p}{\tau_R(t,t')}e^{-\xi(t,t')}f_{eq}(t,t') \,,
\end{equation}
the hydrodynamic generator. The hydrodynamic generator is the sum(integral) of all free streaming contributions at position $(x,t)$ of the effective equilibrium distribution $f_{eq}$ from a region within the light cone weighed by a damping factor.

\textbf{Expansion In Terms Of The Damping Factor :} We observe that the damping exponent $\xi'$ satisfies the relations
\begin{align}
    \xi(t,t) = 0 &,&  \dv{\xi(t,t') }{t'} = -\frac{U\cdot p}{p^{0}\tau_R}  \,.
\end{align}
 
 As both $U\cdot p$ and $\tau_R$ are positive, $\xi(t,t') > 0$. $\xi(t,t') $ is a monotonically decreasing function of $t'$ and is therefore invertible. We now look for the inverse map $ t'(\xi)$. We use the relation for invertible functions,
\begin{align}
    \dv{t' }{\xi'}  = \lrb{\dv{\xi }{t'}}^{-1} = -\frac{p^{0}\tau_R}{U\cdot p}\,.
\end{align}
The inverse can be obtained in the form of a Taylor series expansion \eqref{Ap:Repr}
\begin{align}\label{Eq:tinv}
    t'(\xi,t) -t = \sum_{n = 1}^{\infty} \frac{\xi^{n}}{n!}\lrrb{ \dv{}{\xi'}}^{n-1}\lrb{-\frac{p^{0}\tau_R}{p \cdot U }} \Bigg |_{\xi'=0} \,.
\end{align}
Here we note that for analytic functions, the above Taylor expansion can be written in the Lie series form:
\begin{equation}\label{Eq:txiLie}
    t'(\xi,t) - t = e^{\xi\dv{}{\xi'}}(t'(\xi',t) - t) \Bigg |_{\xi'=0} \,.
\end{equation}
In fact, the evolution of the particle trajectory can be written as, 
\begin{align}\label{Eq:xxiLie}
    x^{\mu} - \frac{p^{\mu}}{p^{0}}(t-t') =  e^{\xi\dv{}{\xi'}} (x^{\mu} - \frac{p^{\mu}}{p^{0}}(t-t'))\Bigg |_{\xi'=0} \,.
\end{align}

\textbf{The Gradient Expansion}: The results, Eq.\eqref{Eq:txiLie} and Eq.\eqref{Eq:xxiLie} allows us to do a coordinate transformation $t' \to \xi'$ and write the hydrodynamic generator as
\begin{equation}\label{Eq:GenLieOut}
     f_{G} = \int_{0}^{\xi_0}d\xi e^{-\xi}e^{\xi\dv{}{\xi'}} f_{eq}( x^{\mu} - \frac{p^{\mu}}{p^{0}}(t-t'))\Bigg |_{\xi'=0}  \,.
\end{equation}
Here, we have used Eq.\eqref{Eq:xxiLie} and the commutative property of the Lie series operator with the function, $ e^{\xi\dv{}{\xi'}}f(x) = f( e^{\xi\dv{}{\xi'}} x)$. The integration limit $\xi_0$ is defined as $\xi(t,t_{0})$. We now use the series expansion of exponential and convert the $\xi'$ derivatives back to $t'$ coordinates using the  relations,
\begin{align}
    \dv{}{\xi'} &= -\frac{p^{0}\tau_R}{U\cdot p}\dv{}{t'}\,, \\
    \dv{}{t'}f(t,t')\Bigg|_{t' =t} &= \frac{p^{\mu}\partial_{\mu}}{p_0}  f(x,t),
\end{align}
and imposing the limits $t' = t$ we get \eqref{Ap:Repr}
\begin{equation}
    f_{G} = \int_{0}^{\xi_0} d\xi e^{-\xi}\sum_{0}^{\infty} \frac{(\xi)^n}{n!}\lrrb{-\frac{\tau_R }{ p \cdot U } \mathcal{D}}^{n} f_{eq}(x,t)\,. \label{Eq:IntGrdSer}
\end{equation}
where $\mathcal{D} = p^{\mu}\partial_{\mu}$. In the general case of Eq.\eqref{Eq:BolEq} the operator $\mathcal{D}$ becomes, 
\begin{equation}
    \mathcal{D} = p^{\mu}\partial_{\mu} + \lrb{mK^{\mu}-\Gamma^{\mu}_{\sigma\rho}p^{\sigma}p^{\rho}}\pdv{}{p^{\mu}} \,.
\end{equation}
  Note that Eq.\eqref{Eq:IntGrdSer} is simply the Taylor expansion of $f_{eq}$ in terms of $\xi'$. The roundabout way in which it was arrived at will be justified when the convergence property of the gradient series is analyzed.

It is easily deduced from Eq.\eqref{Eq:IntGrdSer} that in the limit $\xi_0 \to \infty$ ($t \to \infty$) that $f_{G}$(and therefore $f$) reduces to the Borell re-summation  

\begin{equation}
   \lim_{\xi_0 \to \infty} f_{G} = \int_{0}^{\infty} d\xi e^{-\xi}\sum_{n=0}^{\infty} \frac{(\xi)^n}{n!}\lrrb{-\frac{\tau_R }{ p \cdot U } \mathcal{D}}^{n} f_{eq}(x,t)\,, \label{Eq:IntGrdSerBor}
\end{equation}
of the gradient expansion as conjectured in~\cite{McNelis:2020jrn}.

After integrating Eq.\eqref{Eq:IntGrdSer}, we obtain the exact gradient expansion,
\begin{align}\label{Eq:GrdSer}
    f_{G}  &= \sum_{n=0}^{\infty} \lrb{1 - e^{-\xi_0} \sum_{k=0}^{n}\frac{\xi_0^k}{k!}} \lrrb{-\frac{\tau_R }{ p \cdot U } \mathcal{D}}^{n}f_{\text{eq}}(x,t) \,.
\end{align}
The exact generator therefore is the sum of the usual Chapman-Enskog gradient expansion (first term in the right-hand side of Eq.\eqref{Eq:GrdSer}) corresponding to hydrodynamics and exponentially decaying non-hydrodynamic terms(second term in Eq.\eqref{Eq:GrdSer}). An equivalent result was obtained for the Bjorken case till $n = 40$ and for the general $3+1$D case till $n=3$ in~\cite{McNelis:2020jrn}.

We note that in deriving Eq.\eqref{Eq:GrdSer} no assumption was made about the form of $f_{eq}$, $U^{\mu}$ or $\tau_{R}$ other than the positively of $U\cdot p/\tau_{R}$. Therefore the form of the solution Eq.\eqref{Eq:GrdSer} is the property of the structure of the collision kernel in Eq.\eqref{Eq:RtaMikBol}.
 

\textbf{Convergence}: 
The $\xi_0$ dependent coefficient of the gradient expansion in Eq.\eqref{Eq:GrdSer} is related to the lower incomplete gamma function defined as,
\begin{equation}
    \gamma(n+1,\xi_{0}) = n!\lrb{1 - e^{-\xi_0} \sum_{k=0}^{n}\frac{\xi_0^k}{k!}}\,,
\end{equation} 
which grows asymptotically in $n$ as
\begin{equation}\label{Eq:gammaAsymp}
    \gamma(n,\xi) \sim \frac{e^{-\xi}\xi^n}{n} \left[ 1 + O\!\left(\frac{\xi}{n}\right)\right] \,.
\end{equation}
If the gradient term has a growth rate given by,
\begin{equation}
    \lrrb{-\frac{\tau_R }{ p \cdot U } \mathcal{D}}^{n}f_{\text{eq}}(x,t) \sim n! (K)^n\,,
\end{equation}
where $|K| < \infty$ is some constant which is the maximum value taken by the gradient terms, then the $n$th order term in Eq.\eqref{Eq:GrdSer} grows as $\sim \xi^n(K)^n$. The improvement in convergence is immediately obvious as it has been reduced from factorial to geometric divergence. For non-equilibrium initial conditions, at early times $\xi$ is small and $K$ can be large. If the system under consideration relaxes to equilibrium, then at late times as $\xi \to \infty$ and the gradient $K \to 0$. This implies that convergence is guaranteed if the parameters $K$ and $\xi$ satisfy $\xi K< 1$.

We now show that in general, the gradient expansion, Eq.\eqref{Eq:HydGen}, can have a finite radius of convergence if the damping term $\xi(t',t)$ has a convergent Taylor series (analytic).  First we show that Eq.\eqref{Eq:GrdSer} is convergent if the integrand in Eq.\eqref{Eq:IntGrdSer}, 
\begin{align} \label{Eq:LieSer}
   f_{L} =  \sum_{0}^{\infty} \frac{(\xi)^n}{n!}\lrrb{-\frac{\tau_R }{ p \cdot U } \mathcal{D}}^{n} f_{eq}(x,t)\,,
\end{align}
 is convergent. For this, we first prove the inequality, 
\begin{align}
    \lrb{\int_{0}^{\xi_0} d\xi e^{-\xi} (\xi)^n} \leq \xi_{0}^{n}  \,,
\end{align} 

for $n \geq 1$. Consider the integral
\begin{align}
    \int_{0}^{\xi_0} d\xi e^{-\xi} (\xi)^n  - \xi_0^{n} = \int_{0}^{\xi_0} d\xi \, \lrb{ e^{-\xi} \xi -n}\xi^{n-1} \,.
\end{align}
Observe that $\max{(e^{-\xi} \xi)} = 1/e < 1$ and $\xi \in [0,\xi_0] > 0$. Therefore the term in the bracket is always negative for $n \geq 1$, which proves our proposition. As the $n^{\text{th}}$ order term of $f_{G}$ is less than the corresponding term of $f_{L}$, we have
\begin{equation}
    f_{G} \leq f_{L}\,.
\end{equation}

Therefore by comparison test, the series $f_{G}$ is convergent if $f_{L}$ is convergent. Now, if $f_{eq}$, is an analytic function of $x^{\mu}$ then the convergence of Eq.\eqref{Eq:LieSer} can be reduced to the convergence of Eq.\eqref{Eq:txiLie}. This becomes clear from the equation,
\begin{align}
     e^{-\xi \frac{d}{d\xi'}}f_{eq}\lrrb{x^{i},t,(t-t')} =  f_{eq}\lrrb{x^{i},t,e^{-\xi \frac{d}{d\xi'}}(t-t')}\,.
\end{align}
 We now note that the term $e^{-\xi \frac{d}{d\xi'}}(t-t')$ is the series expansion for the inverse of the function $\xi(t,t')$. We conclude from the Lagrange-Burmann inversion theorem~\cite{Whittaker_Watson_1996}, that if the function $\xi(t,t')$ is analytic at $t' = t$, then the inverse function, $t'(\xi,t)$, has a series expansion at $t' = t$ and has a finite radius of convergence \eqref{Ap:Conv}. 

Note that the proof relies on $\xi(t,t')$ being analytic. But as the Boltzmann equation (Eq.\eqref{Eq:RtaMikBol}) is non-linear, the analyticity of $\xi(t,t')$ depends on $f$ itself unless $U^{\mu}$ and $\tau_R$ are given a priori. So if the initial conditions are analytic and the evolution preserves analyticity, then the gradient series will have a finite radius of convergence at $t$.


\textbf{The Non-Hydrodynamic Corrections:} To get better insight into the physical meaning of the non-hydrodynamic terms, we use integration by parts on Eq.\eqref{Eq:HydGen} and obtain the expression
\begin{align}\label{Eq:OPSol2}
    f_{G} = & \sum_{n=0}^{\infty}\lrcb{ \lrrb{-\frac{\tau_R }{ p \cdot U } D}^{n} f_{eq}(x^{i},t) \right. \nonumber\\
    &\left. - e^{-\xi'(t,t')}\lrrb{-\frac{\tau_R(t,t') }{ p \cdot U(t,t') } D}^{n}  f_{eq}( t,t')\Bigg|_{t' = t_0} } \,.
\end{align}
The form of $f_{G}$ given in Eq.\eqref{Eq:OPSol2} and Eq.\eqref{Eq:GrdSer} are equivalent(if convergent), however, the non-hydrodynamic terms are not the same order-by-order. The two forms are obtained by different summation schemes of the same series \eqref{Ap.EqIntpart}.

We see that the non-hydrodynamic terms in Eq.\eqref{Eq:OPSol2} are free streaming contributions carrying not only the initial data but also information about the history of the evolution of the system. The term $\xi_0$ is a time integral over a function of $U^{\mu}$ and $\tau_R$ which themselves evolve with the system and therefore encode information about the history of evolution. An approach to hydrodynamics, i.e. the decay term vanishing, therefore implies that the system dynamics become history-independent. 

Considering the limits of the coefficients to the gradient terms in Eq.\eqref{Eq:GrdSer}, we see that
\begin{align}
  \lim_{\xi_0 \to \infty}  \lrb{1 - e^{-\xi_0} \sum_{k=0}^{n}\frac{\xi_0^k}{k!}} &= 1 \label{Eq:Limxi} \,,\\
  \lim_{n\to\infty}  \lrb{1 - e^{-\xi_0} \sum_{k=0}^{n}\frac{\xi_0^k}{k!}} &= 0 \,. \label{Eq:Limn} 
\end{align}
The first limit, Eq.\eqref{Eq:Limxi}, implies that when the decay term is large or for long time scales, the initial conditions(and therefore the non-hydrodynamic contributions) become less relevant. The second limit Eq.\eqref{Eq:Limn}, then implies that higher-order gradients are initially suppressed and the lower the order, the faster the decay of the non-hydrodynamic terms. The presence of large initial higher-order gradients implies a slower approach to hydrodynamics. Therefore the knowledge of the non-hydrodynamic terms is crucial in determining the exact contribution to the dynamics from the higher-order gradients at early times. 

Even though the exponential damping is the same at every order, the decay time depends on the polynomial (exponential sum function) 
\begin{equation}
    e_{n}(\xi) \equiv \sum_{k=0}^{n}\frac{\xi_0^k}{k!}
\end{equation} 
multiplying the gradient terms in Eq.\eqref{Eq:GrdSer}. We see that the $k$-th derivative of the pre-factor to the non-hydrodynamic term has the property,
\begin{align}
    \dv[k]{}{\xi}\lrrb{e^{-\xi}e_{n}(\xi)}\Bigg|_{\xi =0} = 0 && k< n \,.
\end{align}
Therefore, an increase in $n$ imply a slower decay rate at $\xi_0 =0$ ($t = t_0$) . From Eq.\eqref{Eq:gammaAsymp}, it can be deduced that for large $n$ the decay time goes as $\xi_0 \sim n$.  The coefficients of the gradients can be interpreted to give length scales associated with each gradient. Observe that we can always write,
\begin{equation}
    \exp{-\xi_0} \propto \exp{-\frac{(t-t_0)}{\sigma_{0}(p,t,t_0)}}\,.
\end{equation}
Here $\sigma_{0}(p,t,t_0)$ gives a local length/time scale within which the free-streaming contributions are significant for the zeroth order gradient. When $\tau_R$ is a constant we have $\sigma_{0} \propto \tau_R$. For higher gradients, this length scale is modified by the polynomial prefactor $e(\xi_{0})$ and goes as $\sigma_{n} \sim \sigma_{0}(p,t,t_0)n$.


\textbf{Domain of Validity:}
Hydrodynamics is traditionally considered valid when the evolution of the energy-momentum tensor is accurately captured by a gradient expansion around the local thermal equilibrium or when the gradient corrections to the equilibrium are small. For the distribution function, this implies,
\begin{equation}
    \left|\frac{f-f_{\text{eq}}}{f_{\text{eq}}}\right| <<1\,.
\end{equation}

To derive a general condition for the validity of a gradient expansion under relaxation time approximation, we look at the full solution 
\begin{align}\label{Eq:FullSol}
    f = e^{-\xi_0}\lrrb{f_{0}(x^{i},t,t_0) -f_{\text{G}}^{H}(x^{i},t,t_0) } + f_{\text{G}}^{H}(x^{i},t)  \,,
\end{align}
where we have rewritten Eq.\eqref{Eq:OPSol2} and used the abbreviation $f_{\text{G}}^{\text{H}}$ for the gradient series. We see that Chapman-Enskog expansion $f_{\text{G}}^{\text{H}}(x^{i},t)$ accurately describes the evolution of the distribution function if the first term of the right hand side of Eq.\eqref{Eq:FullSol} is small. This condition is satisfied for large $\xi_0$. In general, $\xi_0$ can grow arbitrarily fast depending on the initial conditions and the symmetries of the system under consideration. Therefore in the relaxation time model, we have the modified condition for the validity of gradient expansion,
\begin{align}\label{Eq:CondHydRel}
   \left| e^{-\xi_0}\frac{\lrrb{f_0(t,t_0) - f_{\text{G}}^{H}(t,t_0)}}{f_{\text{G}}^{\text{H}}(t)} \right| <<1 \,.
\end{align}
The usual condition for equilibrium hydrodynamics is recovered at $\xi_0=0$ when the gradients are small. From the preceding analysis we can conclude that the scope of the hydrodynamics description, at least under the relaxation time scheme, is broader.

 \textbf{Summary And Outlook}:
We obtain the exact gradient expansion for the Boltzmann equation in the relaxation time approximation. We show that this expansion contains the usual Chapman-Enskog gradient series as well as exponentially decaying transient terms. We prove that the exact solution is the Borel-resummation of the gradient expansion and is convergent under suitable assumptions of analyticity.

 It is important to note that the relaxation time approximation is derived from the linearised Boltzmann collision kernel when the distribution function is close to local thermal equilibrium \cite{Cercignani:2014}. Therefore, the qualitative features of the relaxation time model need not accurately represent the dynamics of a realistic system far from equilibrium. A true vindication of the validity of hydrodynamics out of equilibrium can only come from studying the exact collision kernel that incorporates modifications for strongly interacting systems. However, the insights obtained from the relaxation-time model can be valuable in tackling the exact problem.

\section*{Acknowledgements}
RG and VR acknowledge support from the DAE, Govt. of India. RG thanks Ankit K Panda for the insightful discussions and suggestions, and Dr. Palash Dubey and Dr. Amudhan KU for constructive criticism and valuable feedback.

\newpage
\clearpage
\onecolumngrid
\appendix

\section{Formal solution}\label{Ap:ForSOl}
In this section, we derive the Eq.\eqref{Eq:ExtSol}. The Boltzmann equation in the absence of external forces  in the relaxation time approximation is given by,
\begin{equation}
    p^{\mu}\partial_{\mu}{f} = -\frac{U\cdot p}{\tau_R}(f-f_{\text{eq}}).
\end{equation}

The characteristic curves are given by,
\begin{align}
    \dv{x^{\mu}}{s} =p^{\mu} &&\Rightarrow &&x^{\mu} = p^{\mu}s + c^{\mu}\,.
\end{align}
$s$ is the variable that parametrises the characteristic curve and $c^{\mu}$ is the integration constant representing the initial conditions. 

We can then write the Boltzmann equation along the characteristic curve as
\begin{equation}
\label{eq:A3}
    \frac{d f(x^{\mu}(s),p^{\mu})}{ds} +\frac{U(x^{\mu}(s)) \cdot p}{\tau_R(x^{\mu}(s))} f(x^{\mu}(s),p^{\mu}) =  \frac{U(x^{\mu}(s)) \cdot p}{\tau_R(x^{\mu}(s))} f_{eq}(x^{\mu}(s),p^{\mu}) \,.
\end{equation}
This equation is of the form
\begin{equation}
    \dv{f(s)}{s} + \alpha(s) f(s) = g(s),
\end{equation}
and has the solution \cite{Zaitsev:HandbookPDE}
\begin{equation}
    f(s) = e^{-\int_{0}^{s}\alpha(s^{\prime})ds^{\prime}}f(0) + \int_{0}^{s}e^{-\int_{s^{\prime}}^{s}\alpha(s^{\prime\prime})ds^{\prime\prime}}g(s^{\prime})ds^{\prime}.
\end{equation}

We can now write a formal solution to Eq.\eqref{eq:A3} as
\begin{align}
    f(x^{\mu}(s),p^{\mu}) = \exp{-\int_{0}^{s}\frac{U \cdot p}{\tau_R}(x^{\mu}(s^{\prime})) ds^{\prime}}&  f(x^{\mu}(0),p^{\mu}) \nonumber\\
                            &+ \int_{0}^{s} ds^{\prime} \frac{U\cdot p}{\tau_R}(x^{\mu}(s^{\prime}) \exp{-\int_{s^{\prime\prime}}^{s} \frac{U \cdot p}{\tau_R} ds^{\prime\prime}} f_{eq}(x^{\mu}(s^{\prime}),p^{\mu}).
\end{align}

Here $x^{\mu}(0) = c^\mu $.
We have from the equation of the characteristic curves,
\begin{equation}
    c^{\mu} = x^{\mu} - p^{\mu}s\,.
\end{equation}
 Using $x^{0} = t$ as the parameter.
\begin{align}
    t &= p^{0} s +t_{0}  &&\Rightarrow &&  s = \frac{(t-t_{0})}{p^{0}}, \\
    x^{i} &= \frac{p^{i}}{p^{0}}(t-t_{0}) + c^{i} &&\Rightarrow &&  c^{i} = x^{i} - \frac{p^{i}}{p^{0}}(t-t_{0}). 
\end{align}
also,
\begin{equation}
    ds^{\prime} = \frac{dt^{\prime}}{p^{0}}.
\end{equation}
Writing all the parameter dependencies explicitly, we have the formal solution,

\begin{align}
    f(x^{\mu},p^{\mu}) &= \exp{-  \int_{t_0}^{t}\frac{U(x^i - v^{i}(t-t),t^{\prime})\cdot v }{\tau_R(x^i - v^{i}(t-t^{\prime})},t^{\prime}) dt^{\prime}} f_{0}(x^{i} - v^{i}(t-t_{0}),t_{0},p^{\mu}) \nonumber \\
                            &+ \int_{t_0}^{t} dt^{\prime} \frac{U(x^i - v^{i}(t-t^{\prime}),t^{\prime}) \cdot v}{\tau_R(x^i - v^{i}(t-t^{\prime}))} \exp{ -\int_{t^{\prime}}^{t} \frac{U(x^i - v^{i}(t-t^{\prime\prime}),t^{\prime\prime}) \cdot v}{\tau_R(x^i - v^{i}(t-t^{\prime\prime})),t^{\prime\prime})} dt^{\prime\prime}} f_{eq}(x^i - v^{i}(t-t^{\prime}),t^{\prime})
\end{align}
where $v^{\mu} = p^{\mu}/p^{0}$. For brevity, we introduce the notation
\begin{equation}
    h(t,t') = h(x^i - v^{i}(t-t'),t')
\end{equation}

\subsection{Reparametrization of the integration variable}\label{Ap:Repr}
In this section, we derive the results, Eq.\eqref{Eq:tinv},Eq.\eqref{Eq:txiLie},Eq.\eqref{Eq:xxiLie}.
We define the damping factor
\begin{equation}
    \xi(t,t^{\prime}) = \int_{t^{\prime}}^{t}\frac{U\cdot v }{\tau_R}(t,t^{\prime\prime}) dt^{\prime\prime}\,,
\end{equation} 
with 
\begin{equation}\label{AEq:Xit}
    \dv{\xi}{t^{\prime}}(t,t^{\prime}) = -\frac{U\cdot v }{\tau_R}(t,t^{\prime})\,.
\end{equation}
We define the damping function,
\begin{equation}
    D(t,t^{\prime}) = \exp{-  \xi^{\prime}}\,.
\end{equation}
with the property 
\begin{equation}
    \frac{d}{dt^{\prime}}D(t,t^{\prime}) = \frac{U \cdot v}{\tau_R}D(t,t^{\prime})\,.
\end{equation}

We intend to get the time variable $t'$ as a function of the damping factor $\xi$. For this we Taylor expand $t^{\prime}$ about $t$ or $\xi=0$,
\begin{align}
    t'(t,\xi) &= e^{\xi\dv{}{\xi^{\prime}}} t'(t,\xi^{\prime})\Bigg |_{\xi^{\prime} = 0}\\
               &= t + \sum_{n=1}^{\infty} \frac{(\xi)^n}{n!}\frac{\text{d}^{n}}{\text{d}^{n}\xi^{\prime}}t'(t,\xi^{\prime}) \Bigg |_{\xi^{\prime} = 0}\,.
\end{align}
We now use the relation for invertible functions,
\begin{equation}
    \dv{t' }{\xi}  = \lrb{\dv{\xi }{t'}}^{-1} = -\frac{p^{0}\tau_R}{U\cdot p}\,,
\end{equation}
and get,
\begin{equation}
    \dv[n]{}{\xi}t'(t,\xi) = \dv[n-1]{}{\xi}\lrb{-\frac{p^{0}\tau_R}{U\cdot p}}\,.
\end{equation}
So,
\begin{equation}
    t'(t,\xi) = t + \sum_{n=1}^{\infty} \frac{(\xi)^n}{n!}\frac{\text{d}^{n-1}}{\text{d}^{n-1}\xi^{\prime}}\lrb{-\frac{p^{0}\tau_R}{U\cdot p}}\,.
\end{equation}

As $x^{\mu}$ is independent of $\xi$, $\dv{x^\mu}{\xi} = 0$. This implies,
\begin{equation}
    e^{\xi\dv{}{\xi^{\prime}}}\lrb{x^{\mu} }= x^{\mu}\,.
\end{equation}
So we can write,
\begin{align}\label{AEq:xxiLie}
    x^{\mu} - \frac{p^{\mu}}{p^{0}}(t-t') =  e^{\xi\dv{}{\xi^{\prime}}} (x^{\mu} - \frac{p^{\mu}}{p^{0}}(t-t')(\xi^{\prime}))\Bigg |_{\xi^{\prime}=0} \,.
\end{align}

\subsection{The Gradient Expansion}\label{Ap:GrExp}

Consider the hydrodynamic generator \eqref{Eq:HydGen} 
\begin{equation}
     f_{G} = \int_{0}^{\xi_0}d\xi f_{eq}( x^{\mu} - \frac{p^{\mu}}{p^{0}}(t-t'(\xi)))  \,,
\end{equation}
From Eq.\eqref{AEq:xxiLie} we can write,
\begin{align}
    f_{eq}( x^{\mu} - \frac{p^{\mu}}{p^{0}}(t-t'(\xi)))  &= f_{eq}( e^{\xi\dv{}{\xi^{\prime}}} (x^{\mu} - \frac{p^{\mu}}{p^{0}}(t-t'))\Bigg |_{\xi^{\prime}=0}) \nonumber\\
    &=e^{\xi\dv{}{\xi^{\prime}}} f_{eq}(  (x^{\mu} - \frac{p^{\mu}}{p^{0}}(t-t')))\Bigg |_{\xi^{\prime}=0}\nonumber\\
    & = \sum_{n=0}^{\infty}\frac{(\xi)^n}{n!}\frac{\text{d}^{n}}{\text{d}^{n}\xi^{\prime}}f_{eq}(  (x^{\mu} - \frac{p^{\mu}}{p^{0}}(t-t')))\Bigg |_{\xi^{\prime}=0}\,,
\end{align}
where we have used the commutativity of the Lie series $ e^{\xi\dv{}{\xi^{\prime}}}f(x) = f( e^{\xi\dv{}{\xi^{\prime}}} x)$. The commutativity property of the lie series is equivalent to saying that the Taylor series expansion of the composition of two analytic functions is equal to the composition of their individual Taylor series.

\subsection{Coordinate transformation and the derivatives}
From the chain rule, and Eq.\eqref{AEq:Xit} we have,
\begin{equation}
     \dv{}{\xi} = -\frac{p^{0}\tau_R}{U\cdot p}\dv{}{t^{\prime}} \label{AEq:ChainXit}\,.
\end{equation}
Now consider the function $h \equiv h(x^i-v^i(t-t^{\prime}),t - (t-t^{\prime}))$, its derivative
\begin{align}
     \dv{h}{t'} &\equiv \pdv{h}{x^{i}} \dv{(x^i-v^i(t-t'))}{t'} + \pdv{h}{t} \dv{(t-(t-t^{\prime}))}{t^{\prime}} \nonumber\\
       &= v^i\pdv{h}{x^{i}} + \pdv{h}{t} \,.
\end{align}
At $t^{\prime}= t$ this reduces to 
\begin{align}
     \dv{h}{t'}\Bigg|_{t'=t}  &= v^i\pdv{h}{x^{i}} + \pdv{h}{t}\Bigg|_{t^{\prime}=t} \nonumber\\
                &= \frac{p^{\mu}\partial_{\mu}}{p^{0}} h(x^i,t) \label{AEq:Tdt}\,.
\end{align}

From Eq.\eqref{AEq:ChainXit} and Eq.\eqref{AEq:Tdt}, we have
\begin{equation}
    \dv[n]{}{\xi} = \lrrb{-\frac{\tau_R }{ p \cdot U } p^{\mu}\partial_{\mu}}^{n}.
\end{equation}

This gives
\begin{align}
    f_{eq}( x^{\mu} - \frac{p^{\mu}}{p^{0}}(t-t'(\xi')))   & = \sum_{n=0}^{\infty}\frac{(\xi)^n}{n!}\frac{\text{d}^{n}}{\text{d}^{n}\xi^{\prime}}f_{eq}(  (x^{\mu} - \frac{p^{\mu}}{p^{0}}(t-t')))\Bigg |_{\xi=0}  \nonumber\\
    &= \sum_{n=0}^{\infty}\frac{(\xi)^n}{n!}\lrrb{-\frac{\tau_R }{ p \cdot U } p^{\mu}\partial_{\mu}}^{n}f_{eq}( x^{\mu} ).
\end{align}

\section{Solution from integration by parts}
We can obtain a series expansion of the hydrodynamic generator $f_{G}$ (Eq.\eqref{Eq:HydGen}) in $\xi_0$ by using integration by parts of
\begin{equation*}
     f_{G} = \int_{0}^{\xi_0} d\xi e^{-\xi} f_{eq}(\xi) \,.
\end{equation*}
The first iteration of integration by parts gives,
\begin{align*}
    f_{G} &=    (-1)e^{-\xi} f_{eq}\Bigg |_{0}^{\xi_0} - \int_{0}^{\xi_0} d \xi \frac{d f_{eq} }{d\xi} (-1)e^{-\xi}\,.
\end{align*}

Continuing the integration by parts indefinitely we get the series,
\begin{align}\label{AEq:IBP}
    f_{G} &= \sum_{n = 0}^{\infty} e^{-\xi} (-1)\frac{d^n}{d \xi^n}  \lrb{f_{eq}} \Bigg |_{0}^{\xi_0} \nonumber  \\
    &= \sum_{n = 0}^{\infty} \lrb{ \frac{d^n}{d \xi^n}f_{eq}(0) -  e^{-\xi_0} \frac{d^n}{d \xi^n}  f_{eq}(\xi_0)}.
\end{align}
In obtaining this we assume that,
\begin{equation}
    \lim_{n\to \infty}\int_{0}^{\xi_0} d \xi \frac{\text{d}^n f_{eq} }{\text{d}^n\xi} e^{-\xi} = 0
\end{equation}

\subsection{Equivalace With the Taylor Series Expansion}\label{Ap.EqIntpart}
In this section we show the equivalence between Eq.\eqref{Eq:GrdSer} and Eq.\eqref{Eq:OPSol2}. To show the equivalence of the integration by parts result with the Taylor series expansion, consider the second term in Eq.\eqref{AEq:IBP} and its Taylor expansion about $\xi = 0$. Let
\begin{equation}
    g_{n}(\xi) = \frac{d^n}{d \xi'^n}  f_{eq}(\xi)
\end{equation}That is
\begin{align}
   g_{n}(\xi_0) &= \sum_{m=0}^{\infty} \frac{(\xi_0)^{m}}{m!} \dv[m]{}{\xi} \lrrb{g_{n}(\xi)  } \Bigg|_{\xi = 0} \nonumber\\
                &= \sum_{m=0}^{\infty} \frac{(\xi_0)^{m}}{m!} \dv[m]{}{\xi} \lrrb{\frac{d^n}{d \xi'^n}  f_{eq}(\xi)  } \Bigg|_{\xi = 0} \nonumber \\
                &= \sum_{m=0}^{\infty} \frac{(\xi_0)^{m}}{m!} \dv[m+n]{}{\xi}f_{eq}(\xi) \Bigg|_{\xi = 0}. 
\end{align}
The sum over all $n$ of $g_n$ gives
\begin{align}
    \sum_{n=0}^{\infty} g_{n}(\xi_0) &= \sum_{n=0}^{\infty} \sum_{m=0}^{\infty} \frac{(\xi_0)^{m}}{m!} \dv[m+n]{}{\xi}f_{eq}(\xi) \Bigg|_{\xi = 0}. \nonumber\\
\end{align}
We now re-parametrise the sum in terms of $m+n$ as
\begin{equation}
    \sum_{n=0}^{\infty} g_{n}(\xi_0) = \sum_{l = m+n = 0}^{\infty} e_{l}(\xi_0)\dv[l]{}{\xi}f_{eq}(\xi) \Bigg|_{\xi = 0}
\end{equation}
Where $e_{l}(\xi_0)$ is the exponential sum function, a polynomial that is the sum of all the coefficients of the derivatives that had $m+n = l$. This polynomial is given by
\begin{equation}
    e_{l}(\xi) = \sum_{m = 0}^{l}\frac{\xi^{m}}{m!}.
\end{equation}
We can see this if we write down the sum as
\begin{align}
    \sum_{n=0}^{\infty} g_{n}(\xi_0) &= \sum_{l = m+n = 0}^{\infty} e_{l}(\xi_0)\dv[l]{}{\xi}f_{eq}(\xi) \Bigg|_{\xi = 0} \nonumber\\
    & = f_{eq}(0) \nonumber\\
    & + \dv[1]{}{\xi}f_{eq}(0) + \xi_{0}\dv[1]{}{\xi}f_{eq}(0) \nonumber\\
    & + \dv[2]{}{\xi}f_{eq}(0) + \xi_{0}\dv[2]{}{\xi}f_{eq}(0) + \frac{\xi_{0}^2}{2!}\dv[2]{}{\xi}f_{eq}(0) \nonumber\\
    &  \vdots
\end{align}

\begin{align}
    \sum_{n=0}^{\infty} g_{n}(\xi_0) &= \sum_{l = 0}^{\infty} \lrb{\sum_{k = 0}^{l}\frac{\xi^{k}}{k!}}\dv[l]{}{\xi}f_{eq}(\xi) \Bigg|_{\xi = 0}\nonumber\\
    &= \sum_{l = 0}^{\infty} \lrb{\sum_{k = 0}^{l}\frac{\xi^{k}}{k!}}\dv[l]{}{\xi}f_{eq}(0).
\end{align}
We get the desired result by renaming the dummy summation index $l \Rightarrow n $ and substituting it in Eq.\eqref{AEq:IBP}.

\section{Analyticity and convergence}\label{Ap:Conv}

We define the function
\begin{equation}
    g(t,\Vec{x}) = \frac{U\cdot p}{\tau_R p^0}(t,\Vec{x})\,,
\end{equation}
where $U\cdot p$, $\tau_R$ are functions from $\mathbb{R}^{3+1} \to \mathbb{R}$ and $p^{\mu}$ is a  parameter ($p^{\mu}p_{\mu} = m^2$). As $U\cdot p$ (Energy of the particle in the fluid rest frame) is non-zero and $\tau_R$ (the relaxation time) is finite, the function $g$ is non-zero for all $(t,\Vec{x})$.
\begin{equation}
    \gamma[t'] = (t,\Vec{x}) -(p^0,\Vec{p})\frac{{(t-t')}}{p^0},
\end{equation}
is a curve from $t'\in [t_0,t] \to \mathbb{R}^{3+1}$. Define,
\begin{equation}
    \xi(t') = \int_{t'}^{t}g(\gamma[t''])dt''\,.
\end{equation}
If $g$ is an analytic function at $(t,\Vec{x})$, then $\xi$ is an analytic function of $(t-t')$ at $t-t' = 0$. Then by Lagrange–Burmann inversion theorem, we have an inverse
\begin{equation}
    (t-t'(\xi)) = \sum_{n=1}^{\infty} \frac{a_n}{n!}\xi^n\,
\end{equation}
where
\begin{equation}
    a_n = \lim_{\xi\to 0}\frac{\text{d}^n}{\text{d}t'^n}\lrrb{\frac{(t'-t)}{\xi(t')-\xi(t)}}
\end{equation}
and the series is analytic in the neighbourhood of $\xi = 0$. By uniqueness of power series expansion, one can conclude that this is the same as the series obtained through a Taylor series expansion around $\xi = 0$. 

The composition of analytic functions is analytic. Therefore, if $f(t,\Vec{x})$ is a function from $\mathbb{R}^{3+1} \to \mathbb{R}$ and is analytic at $(t,\Vec{x})$, then $f(\gamma(t')) = f(\gamma(t'(\xi)))$ has a series expansion in $\xi$ that is analytic at $\xi = 0 = (t-t')$.

\bibliography{ref2}

\end{document}